\documentclass[12pt]{article}
\usepackage[dvips]{graphicx}
\usepackage{amsmath}
\usepackage{amsfonts}
\usepackage{amssymb}

\begin{document}

\author{S. Gov\thanks{Also with the Center for Technological Education Holon, 52
Golomb St., P.O.B 305, Holon 58102, Israel.} and S. Shtrikman\thanks{Also with
the Department of Physics, University of California, San Diego, La Jolla,
92093 CA, USA.}\\The Department of Electronics, \\Weizmann Institute of Science,\\Rehovot 76100, Israel
\and H. Thomas\\The Department of Physics and Astronomy,\\University of Basel,\\CH-4056 Basel, Switzerland}
\title{Magnetic Trapping of Neutral Particles: A Study of a Physically Realistic
Model. }
\date{}
\maketitle
\begin{abstract}
Recently, we developed a method for calculating the \emph{lifetime} of the
particle in the special situation where there is no potential barrier, as a
first step in our efforts to understand the quantum-mechanics of magnetic
traps. The toy model that was used in this study was \emph{physically}
\emph{unrealistic} because the magnetic field did not obey Laplace's equation.
Here, we study, both classically and quantum-mechanically, the problem of a
neutral particle with spin $S$, mass $m$ and magnetic moment $\mu$, moving in
two-dimensions in an inhomogeneous physically \emph{realistic} magnetic field
given by
\[
\mathbf{B}=B_{\perp}^{\prime}(x\mathbf{\hat{x}}-y\mathbf{\hat{y}}%
)+B_{0}\mathbf{\hat{z}.}%
\]
We identify
\[
K\equiv\sqrt{\dfrac{S^{2}\left(  B_{\bot}^{\prime}\right)  ^{2}}{\mu
mB_{0}^{3}}},
\]
which is the ratio between the precessional frequency of the particle and its
vibration frequency, as the relevant parameter of the problem.

Classically, we find that when $\mu$ is antiparallel to $\mathbf{B}$, the
particle is trapped provided that $K<\sqrt{4/27}$. We also find that viscous
friction, be it translational or precessional, destabilizes the system.

Quantum-mechanically, we study the problem of spin $S=\hbar/2$ particle in the
same field. Treating $K$ as a small parameter for the perturbation from the
adiabatic Hamiltonian, we find that the lifetime $T_{esc}$ of the particle in
its trapped ground-state is
\[
T_{esc}=\dfrac{T_{vib}}{128\pi^{2}}\exp\left[  \dfrac{2}{K}\right]  \text{ ,}%
\]
where $T_{vib}=2\pi\sqrt{mB_{0}/\mu\left(  B_{\bot}^{\prime}\right)  ^{2}}$ is
the classical period of the particle when placed in the adiabatic potential
$V=\mu\left|  \mathbf{B}\right|  $.
\end{abstract}

\section{Introduction.\label{intro}}

\subsection{Magnetic traps for neutral particles.\label{traps}}

Recently there has been rapid progress in techniques for trapping samples of
neutral atoms at elevated densities and extremely low temperatures. The
development of magnetic and optical traps for atoms has proceeded in parallel
in recent years, in order to attain higher densities and lower temperatures
\cite{t1,t2,t3,t4,t6}. We should note here that neutral traps have been around
much longer than their realizations for neutral atoms might suggest, and the
seminal papers for neutral trapping as applied to neutrons and plasmas date
from the sixties and seventies. Many of these papers are referenced by the
authors of Refs.\cite{t1,t2,t3}. In this paper we concentrate on the study of
\emph{magnetic} traps. Such traps exploit the interaction of the magnetic
moment of the atom with the inhomogeneous magnetic field to provide spatial confinement.

Microscopic particles are not the only candidates for magnetic traps. In fact,
a vivid demonstration of trapping large scale objects is the hovering magnetic
top\cite{levitron,ucas,harrigan,patent}. This ingenues magnetic device, which
hovers in mid-air for about 2 minutes, has been studied recently by several
authors \cite{edge,Berry,bounds,simon,dynamic}.

\subsection{Qualitative description.\label{desc}}

The physical mechanism underlying the operation of magnetic traps is the
adiabatic principle. The common way to describe their operation is in terms of
\emph{classical} mechanics: As the particle is released into the trap, its
magnetic moment points antiparallel to the direction of the magnetic field.
While inside the trap, the particle experiences lateral oscillations
$\omega_{vib}$ which are slow compared to its precession $\omega_{prec}$.
Under this condition the spin of the particle may be considered as
experiencing a \emph{slowly} rotating magnetic field. Thus, the spin precesses
around the \emph{local} direction of the magnetic field $\mathbf{B}$
(adiabatic approximation) and, on the average, its magnetic moment
$\mathbf{\mu}$ points \emph{antiparallel} to the local magnetic field lines.
Hence, the magnetic energy, which is normally given by $-\mathbf{\mu}%
\cdot\mathbf{B}$, is now given (for small precession angle) by $\mu\left|
\mathbf{B}\right|  $. Thus, the overall effective potential seen by the
particle is
\begin{equation}
V_{eff}\simeq\mu\left|  \mathbf{B}\right|  . \label{energy}%
\end{equation}
In the adiabatic approximation, the spin degree of freedom is rigidly coupled
to the translational degrees of freedom, and is already incorporated in
Eq.(\ref{energy}). Thus, under the adiabatic approximation, the particle may
be considered as having only translational degrees of freedom. When the
strength of the magnetic field possesses a\emph{\ minimum}, the effective
potential becomes attractive near that minimum and the whole apparatus acts as
a trap.

As mentioned above, the adiabatic approximation holds whenever $\omega
_{prec}\gg\omega_{vib}$. As $\omega_{prec}$ is inversely proportional to the
spin, this inequality can be satisfied provided that the spin of the particle
is small enough. If, on the other hand, the spin of the particle is too large,
it cannot respond fast enough to the changes of the direction of the magnetic
field. In this limit $\omega_{prec}\ll\omega_{vib}$, the spin has to be
considered as fixed in space and, according to Earnshaw's
theorem\cite{earnshaw}, becomes unstable against \emph{translations}. Note
also that $\omega_{prec}$ is proportional to the field $\left|  \mathbf{B}%
\right|  $. To prevent $\omega_{prec}$ of becoming too small, resulting in
spin-flips (Majorana transitions), most magneto-static traps include a
\emph{bias} field, so that the effective potential $V_{eff}$ possesses a
\emph{nonvanishing} minimum.

\subsection{The purpose and structure of this paper.\label{purp}}

The discussion of magnetic traps in the literature is, almost entirely, done
in terms of \emph{classical} mechanics. In microscopic systems, however,
quantum effects become dominant, and in these cases \emph{quantum mechanics}
is suited for the description of the trap \cite{quant}. An even more
interesting issue is the understanding of how the classical and quantum
descriptions of a \emph{given} system are related.

As a first step in our efforts to understand the quantum-mechanics of magnetic
traps, we recently developed a method for calculating the \emph{lifetime} of
the particle in the special situation where there is no potential
barrier\cite{life1d}. The toy model that was used in this study consisted of a
particle with spin, having only a single translational degree of freedom, in
the presence of a 1D inhomogeneous magnetic field. We found that the trapped
state of the particle decays with a lifetime given by $\sim1/\left(  \sqrt
{K}\omega_{vib}\right)  \exp\left(  2/K\right)  $ where $K=\omega_{vib}%
/\omega_{prec}$. The field that was used in this model was not divergenceless,
and in this sense, the model is \emph{unrealistic}. The next step, presented
in this paper, is to study, both classically and quantum-mechanically, the
case of a particle with spin, having \emph{two} translational degree of
freedom, in the presence of a \emph{physically} \emph{realistic }(i.e.
divergenceless) inhomogeneous magnetic field. This model is reminiscent of a
Ioffe-Pritchard trap\cite{t2,ife}, but without the axial translational degree
of freedom. We neglect the effect of interactions between the particles in the
trap and so we analyze the dynamics of a \emph{single} particle inside the trap.

The structure of this paper is as follows: In Sec.(\ref{def}) we start by
defining the system we study, together with useful parameters that will be
used throughout this paper. Next, we carry out a classical analysis of the
problem in Sec.(\ref{class}). Here, we find two stationary solutions for the
particle inside the trap. One of them corresponds to a state whose spin is
\emph{parallel} to the direction of the magnetic field whereas the other one
corresponds to a state whose spin is \emph{antiparallel }to that direction.
When considering the dynamical stability of these solutions, we find that only
the \emph{antiparallel }stationary solution is stable. We also study the same
problem but with viscous friction added, and arrive at the result that
friction \emph{destabilizes} the system. In Sec.(\ref{quant}) we reconsider
the problem, from a quantum-mechanical point of view. Here, we also find
states that refer to \emph{parallel} and \emph{antiparallel} orientations of
the spin, one of them being bound while the other one unbounded. In this case,
however, these two states are \emph{coupled} due to the inhomogeneity of the
field and we move on to calculate the \emph{lifetime} of the bound state.
Finally, in Sec.(\ref{dis}) we compare the results of the classical analysis
with these of the quantum analysis and comment on their implications to
practical magnetic traps.

\section{Description of the problem.\label{def}}

We consider a particle of mass $m$, magnetic moment $\mu$ and intrinsic spin
$S$ (aligned with $\mu$) moving in an inhomogeneous magnetic field
$\mathbf{B}$ given by
\begin{equation}
\mathbf{B=}B_{0}\mathbf{\hat{z}+}B_{\bot}^{\prime}\left(  x\mathbf{\hat{x}%
-}y\mathbf{\hat{y}}\right)  \text{.} \label{d0}%
\end{equation}
This field possesses a nonzero minimum of amplitude at the origin, which is
the essential part of the trap. The Hamiltonian for this system is%
\begin{equation}
H=\dfrac{p^{2}}{2m}-\mathbf{\mu\cdot B} \label{d0.1}%
\end{equation}
where $p$ is the momentum of the particle.

The Hamiltonian is invariant under a group of operations consisting of a
rotation of position space about the $z$-axis by an arbitrary angle $\gamma$
combined with a rotation of spin space about the $S_{z}$-axis by the
\emph{opposite} angle $-\gamma$. Since the generators of these two rotations
are the $z$-components of orbital angular momentum $L_{z}=xp_{y}-yp_{x}$ and
of spin angular momentum $S_{z}$, respectively, this symmetry gives rise to a
constant of motion,
\begin{equation}
\Lambda=L_{z}-S_{z}=\mathrm{const}. \label{d0.2}%
\end{equation}
Since the magnetic field $\mathbf{B}$ does not depend on $z$, the motion along
the $z$-direction is trivial. Therefore, we restrict ourselves to studying the
motion in the $(x,y)$-plane.

We define $\omega_{prec}$ as the precessional frequency of the particle when
it is at the origin $(x=0,y=0)$. Since at that point the magnetic field is
$\mathbf{B=}B_{0}\mathbf{\hat{z}}$ we find that
\begin{equation}
\omega_{prec}\equiv\dfrac{\mu B_{0}}{S}\text{.} \label{d1}%
\end{equation}
Next, we define $\omega_{vib}$ as the small-amplitude vibrational frequency of
the particle when it is placed in the adiabatic potential field given by
\[
V(x)=\mu\left|  \mathbf{B}(x)\right|  =\mu B_{0}\left(  1+\dfrac{1}{2}\left(
\dfrac{B_{\bot}^{\prime}}{B_{0}}\right)  ^{2}\left(  x^{2}+y^{2}\right)
\right)  +\mathcal{O}\left(  x^{4},x^{2}y^{2},y^{4}\right)  .
\]
For this potential we have
\[
k_{x}=k_{y}=\left.  \dfrac{\partial^{2}V}{\partial x^{2}}\right|  _{@x=0}%
=\mu\dfrac{\left(  B_{\bot}^{\prime}\right)  ^{2}}{B_{0}}\text{,}%
\]
and therefore
\begin{equation}
\omega_{vib}\equiv\sqrt{\dfrac{k_{x}}{m}}=\sqrt{\dfrac{\left(  B_{\bot
}^{\prime}\right)  ^{2}\mu}{mB_{0}}}\text{.} \label{d2}%
\end{equation}
We also define the ratio between $\omega_{vib}$ and $\omega_{prec}$,
\begin{equation}
K\equiv\dfrac{\omega_{vib}}{\omega_{prec}}=\sqrt{\dfrac{S^{2}(B_{\bot}%
^{\prime})^{2}}{\mu mB_{0}^{3}}}\text{.} \label{d3}%
\end{equation}
This will be our `measure of adiabaticity'. It is clear that as $K$ becomes
smaller and smaller, the adiabatic approximation becomes more and more
accurate. Note that when the bias field $B_{0}$ vanishes, $K$ becomes
infinite, and the adiabatic approximation fails. We will later show that,
under this condition, the system become \emph{unstable }against spin flips,
which is in agreement with our discussion at the beginning. This shows that
the introduction of the bias field $B_{0}$, is \emph{essential }to the
operation of the trap with regard to spin-flips. Note also that $K$ is the
only possibility to form a non-dimensional quantity (up to an arbitrary power)
out of the parameters of the system. The value of $K$ therefore, completely
determines the behavior of the system.

\section{Classical analysis.\label{class}}

\subsection{The stationary solutions.\label{stat}}

We denote by $\mathbf{\hat{n}}$ a unit vector in the direction of the spin
(and the magnetic moment). Thus, the equations of motion for the center of
mass of the particle are
\begin{align}
m\dfrac{d^{2}x}{dt^{2}}  &  =\mu\dfrac{\partial}{\partial x}\left(
\mathbf{\hat{n}\cdot B}\right) \label{c1}\\
m\dfrac{d^{2}y}{dt^{2}}  &  =\mu\dfrac{\partial}{\partial y}\left(
\mathbf{\hat{n}\cdot B}\right)  ,\nonumber
\end{align}
and the evolution of its spin is determined by
\begin{equation}
S\dfrac{d\mathbf{\hat{n}}}{dt}=\mu\mathbf{\hat{n}\times B}\text{.} \label{c2}%
\end{equation}
It is straightforward to check that the quantity $\Lambda=L_{z}-S_{z}$ is
indeed conserved.

The two equilibria solutions to Eqs.(\ref{c1}) and (\ref{c2}) are
\begin{equation}
\mathbf{\hat{n}}(t)=\mp\mathbf{\hat{z}} \label{c3}%
\end{equation}
with%
\begin{align*}
x(t)  &  =0\\
y(t)  &  =0
\end{align*}
representing a motionless particle at the origin with its magnetic moment (and
spin) pointing \emph{antiparallel} ($\mathbf{\hat{n}}(t)=-\mathbf{\hat{z}}$)
to the direction of the field at that point and a similar solution but with
the magnetic moment pointing \emph{parallel} to the direction of the field
($\mathbf{\hat{n}}(t)=+\mathbf{\hat{z}}$).

\subsection{Stability of the solutions.}

To check the stability of these solutions we now add first-order
perturbations. We set
\begin{align}
\mathbf{\hat{n}(}t\mathbf{)}  &  =\mathbf{\mp\hat{z}+}\epsilon_{x}%
(t)\mathbf{\hat{x}+}\epsilon_{y}(t)\mathbf{\hat{y}}\label{c4}\\
x(t)  &  =0+\delta x(t)\nonumber\\
y(t)  &  =0+\delta y(t),\nonumber
\end{align}
(note that, to first order, the perturbation $\delta\mathbf{\hat{n}}%
=\epsilon_{x}(t)\mathbf{\hat{x}+}\epsilon_{y}(t)\mathbf{\hat{y}}$ is taken to
be \emph{orthogonal} to the value of $\mathbf{\hat{n}}$ for the stationary
solution $\mathbf{\hat{n}}_{0}=\mathbf{\mp\hat{z}}$, since $\mathbf{\hat{n}}$
is a unit vector) substitute these in Eqs.(\ref{c1}) and (\ref{c2}), and
retain only first-order terms. We find that the resulting equations for
$\delta x(t)$, $\delta y(t)$, $\epsilon_{x}(t)$ and $\epsilon_{y}(t)$ are
\begin{align}
\dfrac{d^{2}\delta x}{dt^{2}}  &  =\dfrac{\mu B_{\bot}^{\prime}}{m}%
\epsilon_{x}\label{c5}\\
\dfrac{d^{2}\delta y}{dt^{2}}  &  =-\dfrac{\mu B_{\bot}^{\prime}}{m}%
\epsilon_{y}\nonumber\\
\dfrac{d\epsilon_{x}}{dt}  &  =\dfrac{\mu}{S}\left(  \mp B_{\bot}^{\prime
}\delta y+B_{0}\epsilon_{y}\right) \nonumber\\
\dfrac{d\epsilon_{y}}{dt}  &  =\dfrac{\mu}{S}\left(  \mp B_{\bot}^{\prime
}\delta x-B_{0}\epsilon_{x}\right)  \text{.}\nonumber
\end{align}
The normal modes of the system transform as the irreducible representations of
the symmetry group. The 4-dimensional linear space spanned by the deviations
$(\delta x,\delta y,\epsilon_{x},\epsilon_{y})$ from the stationary state
carries the irreducible representations $\Gamma_{+}$ with characters
$e^{-i\gamma}$ and $\Gamma_{-}$ with characters $e^{+i\gamma}$, and may thus
be decomposed into the two 2-dimensional invariant subspaces transforming as
$\Gamma_{+}$ and $\Gamma_{-}$, respectively. These subspaces are spanned by
the circular position coordinates and precessional spin coordinates%
\begin{align}
\Gamma_{+}:\;  &  (\rho_{+}=\delta x+i\delta y,\epsilon_{-}=\epsilon
_{x}-i\epsilon_{y});\label{c5.1}\\
\Gamma_{-}:\;  &  (\rho_{-}=\delta x-i\delta y,\epsilon_{+}=\epsilon
_{x}+i\epsilon_{y}). \label{c5.2}%
\end{align}
Thus, the normal modes consist of a circular motion in the $(x,y)$-plane
coupled to a precession of the spin vector in the \emph{opposite} sense.

Indeed, after introducing the $(\rho_{\pm},\epsilon_{\mp})$-coordinates into
Eqs.(\ref{c5}), this set of four equations decomposes into one pair of
equations for $(\rho_{+},\epsilon_{-})$ and another pair for $(\rho
_{-},\epsilon_{+})$. We now look for oscillatory (stable) solutions of these
equations and set
\begin{equation}
\rho_{\pm}=\rho_{\pm,0}e^{-i\omega t},\quad\epsilon_{\pm}=\epsilon_{\pm
,0}e^{-i\omega t}. \label{c7}%
\end{equation}
This yields the algebraic equations
\begin{equation}
\Gamma_{+}:\;\left(
\begin{array}
[c]{cc}%
\omega^{2} & \omega_{vib}^{2}B_{0}/B_{\bot}^{\prime}\\
\pm i\omega_{prec}B_{\bot}^{\prime}/B_{0} & i(\omega+\omega_{prec})
\end{array}
\right)  \cdot\left(
\begin{array}
[c]{l}%
\rho_{+,0}\\
\epsilon_{-,0}%
\end{array}
\right)  =\left(
\begin{array}
[c]{l}%
0\\
0
\end{array}
\right)  , \label{c8.1}%
\end{equation}
$\allowbreak$%
\begin{equation}
\Gamma_{-}:\;\left(
\begin{array}
[c]{cc}%
\omega^{2} & \omega_{vib}^{2}B_{0}/B_{\bot}^{\prime}\\
\mp i\omega_{prec}B_{\bot}^{\prime}/B_{0} & i(\omega-\omega_{prec})
\end{array}
\right)  \cdot\left(
\begin{array}
[c]{l}%
\rho_{-,0}\\
\epsilon_{+,0}%
\end{array}
\right)  =\left(
\begin{array}
[c]{l}%
0\\
0
\end{array}
\right)  . \label{c8.2}%
\end{equation}
These equations have non-trivial solutions whenever the determinant of either
of the two matrices vanishes. This yields the secular equations%

\begin{align}
\Gamma_{+}  &  :\;K\left(  \frac{\omega}{\omega_{vib}}\right)  ^{3}+\left(
\frac{\omega}{\omega_{vib}}\right)  ^{2}\mp1=0,\label{c11.1}\\
\Gamma_{-}  &  :\;K\left(  \frac{\omega}{\omega_{vib}}\right)  ^{3}-\left(
\frac{\omega}{\omega_{vib}}\right)  ^{2}\pm1=0, \label{c11.2}%
\end{align}
which determine the eigenfrequencies $\omega$ of the various modes. Since the
system has three degrees of freedom, we expect to have three normal modes.
Indeed, when $\omega$ is a solution of the first equation, then $-\omega$ is a
solution of the second equation. We define the mode frequencies in
Eq.(\ref{c11.1}) to be positive (or, in the case of complex $\omega$, to have
positive real part); the negative $\omega$-values are needed to construct real
solutions. Then, the $\Gamma_{+}$-modes describe vibrational motions turning
counter-clockwise coupled to spin precessions turning clockwise, i.e.,
opposite to the natural spin precession, and the $\Gamma_{-}$-modes describe
vibrational motions turning clockwise coupled to spin precessions turning
counter-clockwise, i.e., in the same sense as the natural spin precession.

\begin{figure}
[ptb]
\begin{center}
\includegraphics[
height=7.8542in,
width=6.0874in
]%
{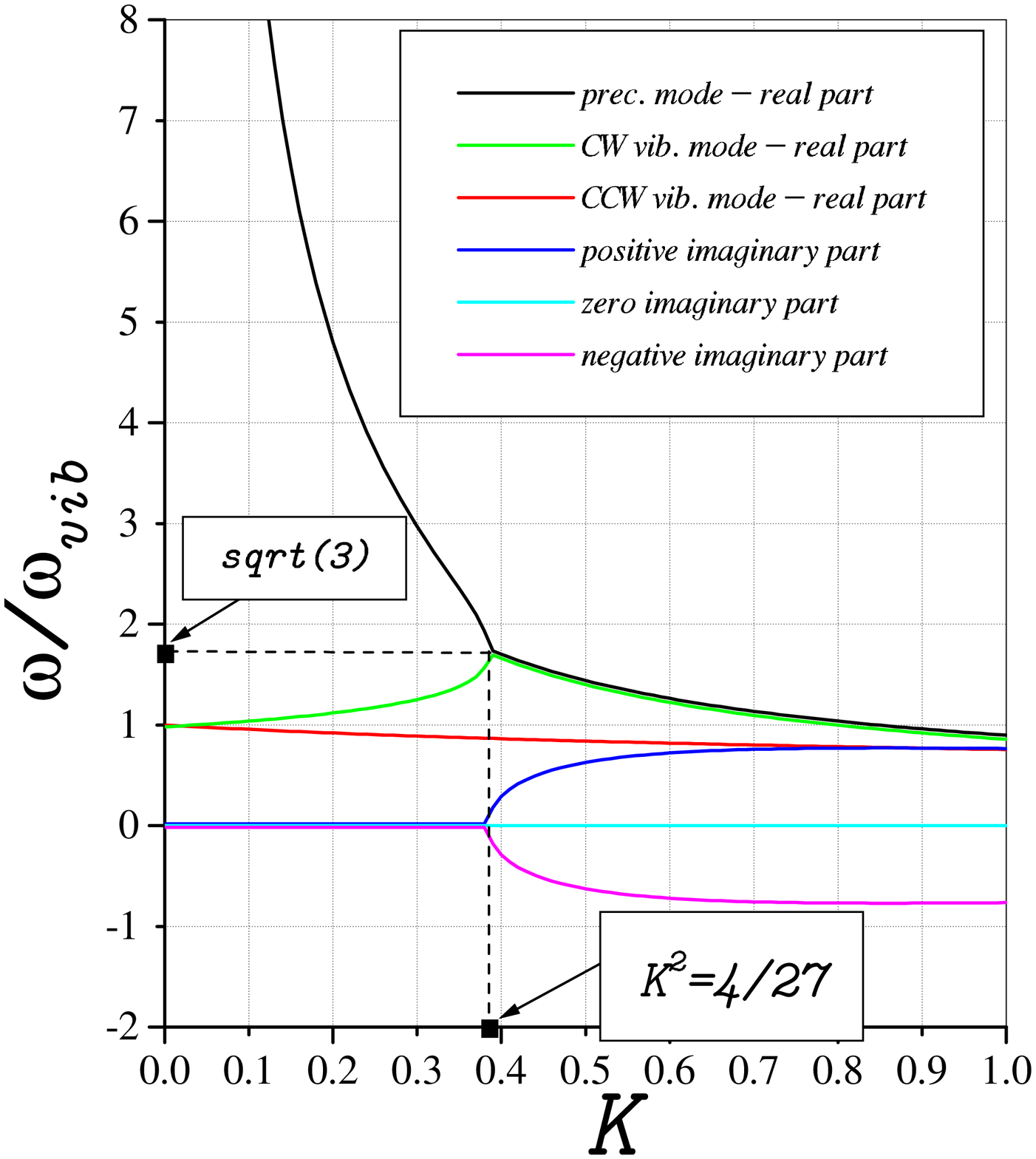}%
\caption{Real and imaginary parts of the mode frequencies as a function of
$K$.}%
\label{fig1}%
\end{center}
\end{figure}

When the \emph{lower} sign is taken in Eqs.(\ref{c4}), corresponding to a spin
\emph{parallel} to the magnetic field, $\mathbf{\hat{n}}=+\mathbf{\hat{z}}$,
we find two $\Gamma_{+}$-modes with complex-conjugate mode frequencies. Thus,
one of the mode frequencies possesses a positive imaginary part, indicating
that the spin-up state is unstable for \emph{any} value of $K$. We therefore
concentrate on the stationary state with the spin \emph{antiparallel} to the
magnetic field, $\mathbf{\hat{n}}=-\mathbf{\hat{z}}$, corresponding to the
\emph{upper} sign in Eqs.(\ref{c4}). For the spin-down state we find one
$\Gamma_{+}$-mode for any value of $K$, and for $K$ smaller than a critical
value $K_{c}=\sqrt{4/27}$ two $\Gamma_{-}$-modes, with real frequencies, as is
shown in Fig.(\ref{fig1}).%

For $K\rightarrow0$, the $\Gamma_{+}$-mode, which goes like $\omega
\simeq\omega_{vib}\left(  1-K/2\right)  $, and the slower of the $\Gamma_{-}%
$-modes, which behaves as $\omega\simeq\omega_{vib}\left(  1+K/2\right)  $,
become degenerate with frequency $\omega=\omega_{vib}$, corresponding to two
linear independent purely vibrational modes. With increasing $K$, the coupling
between translational motion and spin motion lifts the degeneracy and gives
rise to an increasing spin component, which leads to a decrease of the
$\Gamma_{+}$ mode frequency and an increase of the slow $\Gamma_{-}$ mode
frequency. The fastest of the $\Gamma_{-}$-modes, on the other hand, which for
$K\rightarrow0$ is a pure spin precession mode with frequency $\omega
=\omega_{vib}/K=\omega_{prec}$, acquires with increasing $K$ an increasing
vibrational component, leading to a decrease of the mode frequency. At
$K=K_{c}$, it becomes degenerate with the slower $\Gamma_{-}$-mode, with mode
frequency $\omega=\sqrt{3}\omega_{vib}$. For $K>K_{c}$, the two $\Gamma_{-}%
$-modes have complex-conjugate frequencies. Fig.(\ref{fig1}) shows the real
and imaginary parts of the frequencies $\omega$ of the three modes as a
function of $K$.

The dependence of the modes on the parameter $K$ is shown more explicitly by
the form of the eigenvectors. The general form of these is given by%
\begin{equation}
\left(
\begin{array}
[c]{c}%
\rho_{\pm,0}\\
\epsilon_{\mp,0}%
\end{array}
\right)  =\left(
\begin{array}
[c]{c}%
\dfrac{B_{0}}{B_{\perp}^{\prime}}\dfrac{\omega_{vib}}{\omega}\\
-\dfrac{\omega}{\omega_{vib}}%
\end{array}
\right)  A_{\pm}\text{ , } \label{c12.1}%
\end{equation}
where $A_{\pm}$ are dimensionless amplitude parameters.

\subsection{The excitation energy of the modes.}

The excitation energy of a given mode $\xi$ is defined as the difference
between the energy of the mode and the energy of the stationary state,%
\begin{equation}
\xi=-\mu\mathbf{\hat{n}}\cdot\mathbf{B+}\dfrac{1}{2}m\left[  \left(
\dfrac{d\delta x}{dt}\right)  ^{2}+\left(  \dfrac{d\delta y}{dt}\right)
^{2}\right]  -\mu B_{0}. \label{ec1.1}%
\end{equation}
Note that the energy contains bilinear terms in the coordinates and hence, one
cannot neglect the $\hat{z}$-component of the spin. Instead, one must set%
\[
\mathbf{\hat{n}\cdot\hat{z}=-}\sqrt{1-\left(  \epsilon_{x}^{2}+\epsilon
_{y}^{2}\right)  }\mathbf{\simeq}-\left(  1-\dfrac{1}{2}\left(  \epsilon
_{x}^{2}+\epsilon_{y}^{2}\right)  \right)  .
\]
Thus, the correct expression of the energy for small amplitudes is%
\begin{equation}
\xi\simeq-\mu\left[  \dfrac{1}{2}\left(  \epsilon_{x}^{2}+\epsilon_{y}%
^{2}\right)  B_{0}+B_{\perp}^{\prime}\left(  \delta x\epsilon_{x}-\delta
y\epsilon_{y}\right)  \right]  \mathbf{+}\dfrac{1}{2}m\left[  \left(
\dfrac{d\delta x}{dt}\right)  ^{2}+\left(  \dfrac{d\delta y}{dt}\right)
^{2}\right]  . \label{ec1.2}%
\end{equation}
The deviations $\delta x(t),\delta y(t),\epsilon_{x}(t),\epsilon_{y}(t)$ from
the stationary state belonging to the normal modes are given in real form by
\begin{equation}
\delta x(t)=\frac{1}{2}\rho_{\pm,0}e^{-i\omega t}+\mathrm{c.c.},\quad\delta
y(t)=\pm\frac{1}{2i}\rho_{\pm,0}e^{-i\omega t}+\mathrm{c.c.}, \label{ec2}%
\end{equation}%
\begin{equation}
\epsilon_{x}(t)=\frac{1}{2}\epsilon_{\mp,0}e^{-i\omega t}+\mathrm{c.c.}%
,\quad\epsilon_{y}(t)=\mp\frac{1}{2i}\epsilon_{\mp,0}e^{-i\omega
t}+\mathrm{c.c.}. \label{ec3}%
\end{equation}
With the help of Eq.(\ref{c12.1}), one obtains
\begin{equation}
\xi=\mu B_{0}\frac{3\omega_{vib}^{2}-\omega^{2}}{\omega_{vib}^{2}}|A_{\pm
}|^{2}. \label{ec4}%
\end{equation}
From this result we conclude that for $0<K<\sqrt{4/27}$, the excitation energy
of the vibrational modes, for which $\omega^{2}<3\omega_{vib}^{2}$, is
\emph{positive} while the excitation energy of the precessional mode,
satisfying $\omega^{2}>3\omega_{vib}^{2}$, is always \emph{negative}. At the
point $K=\sqrt{4/27}$, where the clockwise vibrational mode and the
precessional mode coalesce, the excitation energy vanishes. We will further
refer to these observations in the following section.

\subsection{The effect of viscous friction.\label{fric}}

When friction is introduced into the system, the equations of motion become
\begin{align}
m\dfrac{d^{2}x}{dt^{2}}  &  =\mu\dfrac{\partial}{\partial x}\left(
\mathbf{\hat{n}\cdot B}\right)  -r_{t}\dfrac{dx}{dt}\label{fric1}\\
m\dfrac{d^{2}y}{dt^{2}}  &  =\mu\dfrac{\partial}{\partial y}\left(
\mathbf{\hat{n}\cdot B}\right)  -r_{t}\dfrac{dy}{dt}\nonumber
\end{align}
and
\begin{equation}
S\dfrac{d\mathbf{\hat{n}}}{dt}=\mu\mathbf{\hat{n}\times B-}r_{p}%
\mathbf{\hat{n}\times}\dfrac{d\mathbf{\hat{n}}}{dt}\text{,} \label{fric2}%
\end{equation}
where $r_{t}$ and $r_{p}$ are translational and precessional friction
coefficients, respectively. The second term in the right-hand side of
Eq.(\ref{fric2}) is the spin-damping contributed by the \emph{change} in the
direction of the spin from $\mathbf{\hat{n}}$ to $\mathbf{\hat{n}%
+}d\mathbf{\hat{n}}$. Since, by definition, $\mathbf{\hat{n}}$ is a unit
vector, it must point perpendicular to $d\mathbf{\hat{n}}$. Thus,
$\Omega_{\perp}=\left|  d\mathbf{\hat{n}/}dt\right|  $ is the angular velocity
associated with the change of $\mathbf{\hat{n}}$. Since the direction of
$\Omega_{\perp}$ must be perpendicular to both $d\mathbf{\hat{n}}$ and
$\mathbf{\hat{n}}$ we form the cross product $\mathbf{\Omega}_{\bot
}=\mathbf{\hat{n}\times}\left(  d\mathbf{\hat{n}/}dt\right)  $ which
incorporates both the correct value and the right direction. Multiplying
$\mathbf{\Omega}_{\bot}$ by $r_{p}$ yields the spin-damping term.

To first order in $r_{r}$ and $r_{t}$ the secular equation in this case is
given by$\allowbreak$
\begin{align}
0  &  =-K^{2}\omega_{n}^{6}+\omega_{n}^{4}-2\omega_{n}^{2}+1+2iK\omega_{n}%
^{5}\dfrac{r_{p}}{S}-2iK^{3}\omega_{n}^{5}\dfrac{r_{t}}{S}\left(  \dfrac
{B_{0}}{B_{\bot}^{\prime}}\right)  ^{2}\label{fric3.1}\\
&  +2iK\omega_{n}^{3}\dfrac{r_{t}}{S}\left(  \dfrac{B_{0}}{B_{\bot}^{\prime}%
}\right)  ^{2}-2iK\omega_{n}^{3}\dfrac{r_{p}}{S}-2iK\omega_{n}\dfrac{r_{t}}%
{S}\left(  \dfrac{B_{0}}{B_{\bot}^{\prime}}\right)  ^{2},\nonumber
\end{align}

where we defined%
\[
\omega_{n}\equiv\dfrac{\omega}{\omega_{vib}}%
\]
to make the expression simple. Let $\omega_{n,0}$ be the eigenfrequencies
$\omega_{n}$ of the frictionless problem, given by Eq.(\ref{c11.2}). When
adding small friction to the problem, the eigenfrequencies will change by a
small amount $\delta\omega_{n}$. We find an approximate expression for
$\delta\omega_{n}$ by expanding Eq.(\ref{fric3.1}) around $\omega_{n,0}$ to
first order in $\delta\omega_{n}$ and making use of Eq.(\ref{c11.2}). This
gives
\begin{equation}
\delta\omega_{n}=\dfrac{iK}{S}\left(  \dfrac{r_{p}\omega_{n,0}^{4}%
+r_{t}\left(  B_{0}/B_{\perp}^{\prime}\right)  ^{2}}{\omega_{n,0}^{2}%
-3}\right)  +\mathcal{O}\left(  r_{t}^{2},r_{t}r_{p},r_{p}^{2}\right)  .
\label{fric4}%
\end{equation}

Eq.(\ref{fric4}) has an interesting consequence: The numerator in
Eq.(\ref{fric4}) is positive for all three modes while the denominator is
negative for the two vibrational modes and positive for the precessional mode.
We therefore conclude that friction, either translational or precessional,
stabilizes the vibrational modes and, simultaneously, destabilizes the
precessional mode. The system all together becomes of course, \emph{unstable}.

The fact that spin damping leads to an exponential growth of the precessional
mode is no surprise in view of its negative excitation energy. Also, the
exponential decay of the vibrational modes due to translational friction is to
be expected on account of their positive excitation energy. What \emph{is}
important is the fact that due to the coupling between translation and
precession, \emph{translational} friction causes an exponential growth of the
\emph{precessional} mode, with a growth time which, compared to the effect of
spin damping, is smaller by a factor of $r_{t}K^{2}S^{2}/\mu mr_{p}B_{0}^{2}$
in the limit of small $K$.

\section{Quantum-mechanical analysis.\label{quant}}

\subsection{The Hamiltonian and its diagonalized form.\label{ham}}

In this section we consider the problem of a neutral particle with spin half (
$S=\hbar/2)$ in a 2D inhomogeneous magnetic field from a quantum-mechanical
point of view. Unlike the classical analysis, in which the derivation was
valid for any value of the adiabaticity parameter $K$, we concentrate here on
the behavior of the system when $K$ is \emph{small}. We choose to analyze the
case of a spin half particle because this case already shows the essentials of
the quantum-mechanical problem. Note also that, quantum mechanically, the
magnetic moment $\mu$ and the spin $S$ of a particle are related by%
\[
\mu=\gamma S,
\]
where $\gamma$ is the gyromagnetic ratio of the particle. Setting $\mu=\gamma
S$ and $S=\hbar/2$ in Eq.(\ref{d3}) gives%
\[
K=\sqrt{\dfrac{\hbar(B_{\bot}^{\prime})^{2}}{2\gamma mB_{0}^{3}}}.
\]

Now, it is convenient to express the spatial dependence of the magnetic field
in polar coordinates $r=\sqrt{x^{2}+y^{2}}$ and $\phi=\arctan\left(
y/x\right)  $. We also denote by $B$ the amplitude of $\mathbf{B}$, by
$\theta$-its direction with respect to the $\mathbf{\hat{z}}$ axis and by
$\varphi$ the angle between the projection of $\mathbf{B}$ onto the $x$-$y$
plane and the $\hat{x}$-axis. Thus, Eq.(\ref{d0}) is rewritten as
\begin{equation}
\mathbf{B}=B\left[  \sin\theta\cos\varphi\mathbf{\hat{x}+}\sin\theta
\sin\varphi\mathbf{\hat{y}}+\cos\theta\mathbf{\hat{z}}\right]  \label{h1}%
\end{equation}
where
\begin{align}
B  &  =B_{0}\sqrt{1+\left(  \dfrac{B_{\bot}^{\prime}}{B_{0}}\right)  ^{2}%
r^{2}}\text{ ,}\label{h1.1}\\
\theta &  =\arctan\left(  \dfrac{B_{\bot}^{\prime}r}{B_{0}}\right)
,\nonumber\\
\varphi &  =\arctan\left(  \dfrac{B_{y}}{B_{x}}\right)  =-\arctan\left(
\dfrac{y}{x}\right)  =-\phi.\nonumber
\end{align}
Thus, $B$ and $\theta$ depends only on $r$ whereas $\varphi$ depends
(linearly) only on $\phi$.

The time-independent Schr\"{o}dinger equation for this system is
\begin{equation}
\left[  -\frac{\hbar^{2}}{2m}\nabla^{2}-\mu B\left(  \sin\theta\cos\varphi
\hat{\sigma}_{x}\mathbf{+}\sin\theta\sin\varphi\hat{\sigma}_{y}+\cos\theta
\hat{\sigma}_{z}\right)  \right]  \Psi^{\prime}\left(  r,\phi\right)
=E\Psi^{\prime}\left(  r,\phi\right)  \label{h5}%
\end{equation}
where $\hat{\sigma}_{x}$, $\hat{\sigma}_{y}$ and $\hat{\sigma}_{z}$ are the
Pauli matrices given by
\[%
\begin{array}
[c]{ccc}%
\hat{\sigma}_{x}=\left(
\begin{array}
[c]{cc}%
0 & 1\\
1 & 0
\end{array}
\right)  & \hat{\sigma}_{y}=\left(
\begin{array}
[c]{cc}%
0 & -i\\
i & 0
\end{array}
\right)  & \hat{\sigma}_{z}=\left(
\begin{array}
[c]{cc}%
1 & 0\\
0 & -1
\end{array}
\right)  ,
\end{array}
\]
$E$ is the eigenenergy and $\Psi^{\prime}$ is the two-components spinor
\begin{equation}
\Psi^{\prime}=\left(
\begin{array}
[c]{c}%
\psi_{\uparrow}^{\prime}\left(  r,\phi\right) \\
\psi_{\downarrow}^{\prime}\left(  r,\phi\right)
\end{array}
\right)  . \label{h5.1}%
\end{equation}
In matrix form Eq.(\ref{h5}) becomes
\begin{equation}
\left(  H_{K}+H_{M}\right)  \left(
\begin{array}
[c]{c}%
\psi_{\uparrow}^{\prime}\left(  r,\phi\right) \\
\psi_{\downarrow}^{\prime}\left(  r,\phi\right)
\end{array}
\right)  =E\left(
\begin{array}
[c]{c}%
\psi_{\uparrow}^{\prime}\left(  r,\phi\right) \\
\psi_{\downarrow}^{\prime}\left(  r,\phi\right)
\end{array}
\right)  \label{h6.0}%
\end{equation}
where $H_{K}$ and $H_{M}$, given by
\begin{align}
H_{K}  &  \equiv-\dfrac{\hbar^{2}}{2m}\nabla^{2}\label{h6.1}\\
H_{M}  &  \equiv-\mu B\left(
\begin{array}
[c]{cc}%
\cos\theta & \sin\theta e^{-i\varphi}\\
\sin\theta e^{i\varphi} & -\cos\theta
\end{array}
\right)  ,\nonumber
\end{align}
are the kinetic part and the magnetic part of the Hamiltonian $H$, respectively.

In order to diagonalize the magnetic part of the Hamiltonian, we make a local
\emph{passive} transformation of coordinates on the wavefunction such that the
spinor is expressed in a new coordinate system whose $\mathbf{\hat{z}}$ axis
coincides with the direction of the magnetic field at the point $\left(
r,\phi\right)  $. We denote by $R\left(  r,\phi\right)  $ the required
transformation and set $\Psi=R\Psi^{\prime}$. Thus, $\Psi$ represent \emph{the
same} direction of the spin as before the transformation but using the
\emph{new} coordinate system. The Hamiltonian in this newly defined system is
clearly given by $RHR^{-1}$. In the case of the magnetic field given in
Eqs.(\ref{h1}) and (\ref{h1.1}) the required transformation is accomplished by
using the three Euler angles: First, we perform a rotation through an angle
$\varphi$ around the $\hat{z}$ axis. Second, we make a rotation through an
angle $\theta$ around the \emph{new} position of the $\hat{y}$ axis. At the
end of this process the new $\hat{z}$ axis coincide with the direction of the
magnetic field. Now the value of the last Euler angle, which is a rotation
around the new $\hat{z}$ axis, has no effect on this axis. For simplicity we
choose this angle to be $0$. Thus, the representation of the complete
transformation for spin half particle is given by \cite{rot}
\[
R=\exp\left[  i\dfrac{\theta}{2}\hat{\sigma}_{y}\right]  \exp\left[
i\dfrac{\varphi}{2}\hat{\sigma}_{z}\right]  ,
\]
while its inverse is given by
\[
R^{-1}=\exp\left[  -i\dfrac{\varphi}{2}\hat{\sigma}_{z}\right]  \exp\left[
-i\dfrac{\theta}{2}\hat{\sigma}_{y}\right]  .
\]
It is easily verified that the transformation indeed diagonalizes the magnetic
part of the Hamiltonian as%

\[
RH_{M}R^{-1}=-\mu B\hat{\sigma}_{z}.
\]
For the kinetic part we find, after some algebra, that
\[
RH_{K}R^{-1}=-\dfrac{\hbar^{2}}{2m}\left[
\begin{array}
[c]{c}%
\mathbf{\nabla}^{2}-\dfrac{1}{4}\left(  \dfrac{d\theta}{dr}\right)
^{2}-\dfrac{1}{4r^{2}}\\
+\dfrac{i}{r^{2}}\hat{\sigma}_{z}\cos\theta\dfrac{\partial}{\partial\phi}\\
-i\left[  \left(  \dfrac{d\theta}{dr}\right)  \dfrac{\partial}{\partial
r}+\dfrac{1}{2}\dfrac{d^{2}\theta}{dr^{2}}+\dfrac{1}{2r}\dfrac{d\theta}%
{dr}\right]  \hat{\sigma}_{y}\\
-\dfrac{i}{r^{2}}\hat{\sigma}_{x}\sin\theta\dfrac{\partial}{\partial\phi}%
\end{array}
\right]
\]

Thus, the Hamiltonian of the system in the rotated frame may be written as
\begin{equation}
H=H_{diag}+H_{int} \label{h7}%
\end{equation}
where
\begin{align}
H_{diag}  &  =-\dfrac{\hbar^{2}}{2m}\left[  \mathbf{\nabla}^{2}-\dfrac{1}%
{4}\left(  \dfrac{d\theta}{dr}\right)  ^{2}-\dfrac{1}{4r^{2}}+\dfrac{i}{r^{2}%
}\hat{\sigma}_{z}\cos\theta\dfrac{\partial}{\partial\phi}\right]  -\mu
B\hat{\sigma}_{z}\label{h7.01}\\
H_{int}  &  =-\dfrac{\hbar^{2}}{2m}\left\{  -i\hat{\sigma}_{y}\left[  \left(
\dfrac{d\theta}{dr}\right)  \dfrac{\partial}{\partial r}+\dfrac{1}{2}%
\dfrac{d^{2}\theta}{dr^{2}}+\dfrac{1}{2r}\dfrac{d\theta}{dr}\right]
-\dfrac{i}{r^{2}}\hat{\sigma}_{x}\sin\theta\dfrac{\partial}{\partial\phi
}\right\}  .\nonumber
\end{align}
The first part of the Hamiltonian $H_{diag}$ is diagonal. It contains the
kinetic part $\sim\mathbf{\nabla}^{2}$, a term whose form is $\mp$ $\mu B$
which is to be identified as the adiabatic effective potential and the terms
$\sim1/r^{2},ir^{-2}\hat{\sigma}_{z}\partial/\partial\phi$ which appear due to
the rotation. The second part of the Hamiltonian $H_{int}$ contains only
non-diagonal components. These will be shown to be of order $\mathcal{O}%
\left(  K\right)  $ and hence may be regarded as a small perturbation. We
proceed to find the eigenstates of $H_{diag}$.

\subsection{Stationary states of $H_{diag}$.\label{diag}}

Since $H_{diag}$ is diagonal, the two spin states of the wavefunction are
decoupled. We then seek a solution of the form
\begin{equation}
\Psi_{\downarrow}=\left(
\begin{array}
[c]{c}%
0\\
\psi_{\downarrow}(r,\phi)
\end{array}
\right)  \text{ ; }E=E_{\downarrow}, \label{h8.01}%
\end{equation}
referred to as the \emph{spin-down} state, and another solution
\begin{equation}
\Psi_{\uparrow}=\left(
\begin{array}
[c]{c}%
\psi_{\uparrow}(r,\phi)\\
0
\end{array}
\right)  \text{ ; }E=E_{\uparrow}, \label{h8.02}%
\end{equation}
which we call the \emph{spin-up} state.

The equation for the non-vanishing component of the spin-down state is given
by
\begin{equation}
\left\{  -\dfrac{\hbar^{2}}{2m}\left[  \mathbf{\nabla}^{2}-\dfrac{1}{4}\left(
\dfrac{d\theta}{dr}\right)  ^{2}-\dfrac{1}{4r^{2}}-\dfrac{i}{r^{2}}\cos
\theta\dfrac{\partial}{\partial\phi}\right]  +\mu B\right\}  \psi_{\downarrow
}=E_{\downarrow}\psi_{\downarrow}, \label{h8.1}%
\end{equation}
whereas the equation for the non-vanishing component of the spin-up state is
\begin{equation}
\left\{  -\dfrac{\hbar^{2}}{2m}\left[  \mathbf{\nabla}^{2}-\dfrac{1}{4}\left(
\dfrac{d\theta}{dr}\right)  ^{2}-\dfrac{1}{4r^{2}}+\dfrac{i}{r^{2}}\cos
\theta\dfrac{\partial}{\partial\phi}\right]  -\mu B\right\}  \psi_{\uparrow
}=E_{\uparrow}\psi_{\uparrow}. \label{h8.2}%
\end{equation}

We now show that in the limit of small $K$ we can neglect the term
$\sim\left(  d\theta/dr\right)  ^{2}$ in both Eq.(\ref{h8.1}) and
Eq.(\ref{h8.2}): We compare the order of magnitude of the term $\mu B$ to that
of the term $\hbar^{2}\left(  d\theta/dr\right)  ^{2}/8m$. Using
Eq.(\ref{h1.1}) it can be easily shown that the maximum value of $d\theta/dr$
is $B_{\bot}^{\prime}/B_{0}$ whereas the minimum value of $\mu B$ is $\mu
B_{0}$. Thus,
\[
\dfrac{\left.  \mu B\right|  _{\text{min}}}{\left(  \dfrac{\hbar^{2}}%
{8m}\left(  \dfrac{d\theta}{dr}\right)  _{\text{max}}^{2}\right)  }%
=\dfrac{8\mu mB_{0}^{3}}{\left(  B_{\bot}^{\prime}\right)  ^{2}\hbar^{2}%
}=\frac{2}{K^{2}},
\]
and hence we can neglect the term $\sim(d\theta/dr)^{2}$ when $K$ is small.
Furthermore, as we are interested in the solutions near the origin we replace
the $\cos\theta$ term by its zeroth-order approximation around $r=0$. We will
justify this approximation later. Under these approximations, Eqs.(\ref{h8.1})
and (\ref{h8.2}) simplify to
\begin{equation}
\left\{  -\dfrac{\hbar^{2}}{2m}\left[  \mathbf{\nabla}^{2}-\dfrac{1}{4r^{2}%
}-\dfrac{i}{r^{2}}\dfrac{\partial}{\partial\phi}\right]  +\mu B\right\}
\psi_{\downarrow}=E_{\downarrow}\psi_{\downarrow} \label{h8.3}%
\end{equation}
and
\begin{equation}
\left\{  -\dfrac{\hbar^{2}}{2m}\left[  \mathbf{\nabla}^{2}-\dfrac{1}{4r^{2}%
}+\dfrac{i}{r^{2}}\dfrac{\partial}{\partial\phi}\right]  -\mu B\right\}
\psi_{\uparrow}=E_{\uparrow}\psi_{\uparrow}. \label{h8.4}%
\end{equation}

The approximate solutions of these equations is outlined in the next two subsections.

\subsubsection{Stationary spin-down states.\label{down}}

Eq.(\ref{h8.3}) represents a particle in a symmetric \emph{attractive}
potential. If the extent of the wave function is small enough we can expand
$B$ in Eq.(\ref{h1.1}) to second order in $r$%
\begin{equation}
B\simeq B_{0}\left[  1+\dfrac{1}{2}\left(  \dfrac{B_{\bot}^{\prime}r}{B_{0}%
}\right)  ^{2}\right]  +\mathcal{O}\left(  r^{4}\right)  , \label{down0}%
\end{equation}
and apply the well-known solution of the harmonic oscillator\cite{sho} in two
dimensions. Under this approximation, Eq.(\ref{h8.3}) becomes%
\begin{equation}
\left[  -\dfrac{\hbar^{2}}{2m}\left(  \dfrac{1}{r}\dfrac{\partial}{\partial
r}+\dfrac{\partial^{2}}{\partial r^{2}}-\dfrac{1}{r^{2}}\left(  i\dfrac
{\partial}{\partial\phi}+\dfrac{1}{2}\right)  ^{2}\right)  +\dfrac{\mu\left(
B_{\bot}^{\prime}\right)  ^{2}r^{2}}{2B_{0}}\right]  \psi_{\downarrow}=\left(
E_{\downarrow}-\mu B_{0}\right)  \psi_{\downarrow}. \label{down0.1}%
\end{equation}
We seek a solution whose form is%
\begin{equation}
\psi_{\downarrow}(r,\phi)=f(r)e^{i\nu\phi} \label{down0.2}%
\end{equation}
and then the equation satisfied by $f\left(  r\right)  $ is%
\begin{equation}
-\dfrac{\hbar^{2}}{2m}\left[  \dfrac{1}{r}\dfrac{df}{dr}+\dfrac{d^{2}f}%
{dr^{2}}-\dfrac{f}{r^{2}}\left(  \nu-\dfrac{1}{2}\right)  ^{2}\right]
+\dfrac{\mu\left(  B_{\perp}^{\prime}\right)  ^{2}r^{2}f}{2B_{0}}=\left(
E_{\downarrow}-\mu B_{0}\right)  f, \label{down0.3}%
\end{equation}
which is an eigenvalue problem for $f$. The smallest eigenvalue for this
problem is obtained by setting%
\[
\nu=\dfrac{1}{2},
\]
for which the eigenfunction $f$ is%
\[
f\left(  r\right)  =D\exp\left[  -\sqrt{\dfrac{\mu m\left(  B_{\perp}^{\prime
}\right)  ^{2}}{4\hbar^{2}B}}r^{2}\right]  =D\exp\left[  -\dfrac{1}{4K}\left(
\dfrac{B_{\perp}^{\prime}r}{B_{0}}\right)  ^{2}\right]  .
\]
Thus, under the harmonic oscillator approximation the down-part of the
spin-down state is
\begin{equation}
\psi_{\downarrow}=\frac{B_{\perp}^{\prime}}{B_{0}\sqrt{2\pi K}}\exp\left[
-\dfrac{1}{4K}\left(  \dfrac{B_{\perp}^{\prime}r}{B_{0}}\right)  ^{2}\right]
e^{i\phi/2}\text{ }. \label{down1}%
\end{equation}
where the normalization constant $D$ has been calculated by demanding that%
\[%
{\displaystyle\int\limits_{0}^{\infty}}
rdr%
{\displaystyle\int\limits_{0}^{2\pi}}
d\phi\left|  \psi_{\downarrow}\right|  ^{2}=1,
\]
using the definite integral%
\[%
{\displaystyle\int\limits_{0}^{\infty}}
re^{-ar^{2}}dr=\dfrac{1}{2a}.
\]
Note that the extent of this wave function over which it changes appreciably
is given by
\begin{equation}
\Delta r_{\downarrow}\sim\sqrt{K}\dfrac{B_{0}}{B_{\bot}^{\prime}},
\label{down2}%
\end{equation}
whereas the extent over which $\mu B$ changes significantly (see
Eq.(\ref{h1.1})) is
\begin{equation}
\Delta r_{\mu B}\sim\dfrac{B_{0}}{B_{\bot}^{\prime}}. \label{down3}%
\end{equation}
Thus, the ratio between these two length scales is
\begin{equation}
\dfrac{\Delta r_{\downarrow}}{\Delta r_{\mu B}}\sim\sqrt{K}. \label{down4}%
\end{equation}
We therefore conclude that when $K$ is small enough, the harmonic
approximation is justified. Note also that $\Delta r_{\mu B}$ is also the
typical length of $\cos\theta$. This shows that the substitution of
$\cos\theta$ in Eq.(\ref{h8.1}) by $1$ is also justified.

The wave function $\psi_{\downarrow}$, given by Eq.(\ref{down1}), then
represents the lowest possible bound state for this system. This state
corresponds to a \emph{trapped} particle. The energy of this state is clearly
\begin{equation}
E_{\downarrow}=\mu B_{0}+2\left(  \frac{\hbar}{2}\omega_{vib}\right)  =\mu
B_{0}\left(  1+2K\right)  \simeq\mu B_{0}, \label{down5}%
\end{equation}
while its full spinor representation is
\begin{equation}
\Psi_{\downarrow}=\left(
\begin{array}
[c]{c}%
0\\
\dfrac{B_{\perp}^{\prime}}{B_{0}\sqrt{2\pi K}}\exp\left[  -\dfrac{1}%
{4K}\left(  \dfrac{B_{\perp}^{\prime}r}{B_{0}}\right)  ^{2}\right]
e^{i\phi/2}%
\end{array}
\right)  . \label{down6}%
\end{equation}

\subsubsection{Stationary spin-up states.\label{up}}

Eq.(\ref{h8.4}) describes a particle in a \emph{repulsive} potential. It
corresponds to an unbounded state representing an \emph{untrapped} particle.
In this case there is a continuum of states, each with its own energy. As we
are interested in non-radiative decay, we focus on finding a solution with an
energy which is \emph{equal} to the energy found for the trapped state, that
is
\begin{equation}
E_{\uparrow}=E_{\downarrow}\simeq\mu B_{0}. \label{up0}%
\end{equation}
When evaluating the lifetime in the next section, we compute the matrix
element of $H_{int}$ between the states $\psi_{\uparrow}$ and $\psi
_{\downarrow}$. Thus, most of the contribution to this integral comes from the
region in $r$ where $\psi_{\downarrow}$ is substantial. According to
Eq.(\ref{down4}), $\mu B$ changes very little in this range and, as a first
approximation, we may take $\cos\theta\simeq1$ and the potential in this
region as \emph{uniform},%
\begin{equation}
\mu B\simeq\mu B_{0} \label{up0.1}%
\end{equation}
in Eq.(\ref{h8.4}). We now set a solution whose form is%
\[
\psi_{\uparrow}(r,\phi)=g(r)e^{i\gamma\phi}.
\]
Substituting this, together with Eqs.(\ref{up0}) and (\ref{up0.1}) into
Eq.(\ref{h8.4}) gives%
\[
-\dfrac{\hbar^{2}}{2m}\left[  \dfrac{1}{r}\dfrac{dg}{dr}+\dfrac{d^{2}g}%
{dr^{2}}-\dfrac{g}{r^{2}}\left(  \gamma+\dfrac{1}{2}\right)  ^{2}\right]
=2\mu B_{0}g,
\]
whose non-singular solution is%
\[
g\left(  r\right)  =J_{\gamma+1/2}\left(  \dfrac{B_{\perp}^{\prime}r}{B_{0}%
K}\right)  ,
\]
where $J_{\alpha}\left(  x\right)  $ is the Bessel function of the first kind
of order $\alpha$.

We note that all the four terms of $H_{int}$ does not operate on the $\phi$
coordinate. This is a consequence of the fact that $L_{z}-S_{z}$ (where
$L_{z}$ is the $z$-component of the orbital angular momentum and $S_{z}%
=\hbar\sigma_{z}/2$ is the $z$-component of the spin) is conserved. Hence, in
order to have a non-vanishing matrix element between the up-state and the
down-state, they must have the \emph{same} $\phi$-dependence. Thus,
$\gamma=\nu=1/2$ and as a result
\begin{equation}
\psi_{\uparrow}=CJ_{1}\left(  \dfrac{B_{\perp}^{\prime}r}{B_{0}K}\right)
e^{i\phi/2} \label{up0.2}%
\end{equation}
with%
\begin{equation}
\Psi_{\uparrow}=\left(
\begin{array}
[c]{c}%
CJ_{1}\left(  \dfrac{B_{\perp}^{\prime}r}{B_{0}K}\right)  e^{i\phi/2}\\
0
\end{array}
\right)  . \label{up0.3}%
\end{equation}
where $C$ is the normalization constant which is chosen to be real.

The wave function given in Eq.(\ref{up0.2}) is oscillatory. It has a period of
about
\begin{equation}
\Delta r_{\uparrow}\sim K\dfrac{B_{0}}{B_{\perp}^{\prime}} \label{up5}%
\end{equation}
near the origin. Comparing it to $\Delta r_{\downarrow}$ given in
Eq.(\ref{down2}), we find that
\begin{equation}
\dfrac{\Delta r_{\uparrow}}{\Delta r_{\downarrow}}\sim\sqrt{K}, \label{up6}%
\end{equation}
which shows that, for $K\ll1$, the wavefunction $\psi_{\uparrow}$ executes
many oscillations in the region where $\psi_{\downarrow}$ is appreciable.

\subsection{The lifetime.\label{time}}

To evaluate the lifetime $T_{esc}$ of the particle in its trapped state, which
is the average time it takes for the particle to escape, we calculate the
transition rate from the bound state given by Eq.(\ref{down6}), to the
unbounded state Eq.(\ref{up0.3}), according to Fermi's golden rule\cite{fermi}%
. Thus,
\begin{equation}
\dfrac{1}{T_{esc}}=\dfrac{2\pi}{\hbar}\left|  H_{\downarrow,\uparrow}\right|
^{2}g(E_{\uparrow}) \label{t1}%
\end{equation}
where
\begin{align}
H_{\downarrow,\uparrow}  &  =%
{\displaystyle\int\limits_{0}^{\infty}}
rdr%
{\displaystyle\int\limits_{0}^{2\pi}}
d\phi\Psi_{\uparrow}^{\dagger}H_{int}\Psi_{\downarrow}\label{t2}\\
&  =-\dfrac{\hbar^{2}}{2m}%
{\displaystyle\int\limits_{0}^{\infty}}
rdr%
{\displaystyle\int\limits_{0}^{2\pi}}
d\phi\psi_{\uparrow}^{\ast}\left\{  -\left(  \dfrac{d\theta}{dr}\right)
\dfrac{\partial}{\partial r}-\dfrac{1}{2}\dfrac{d^{2}\theta}{dr^{2}}-\dfrac
{1}{2r}\dfrac{d\theta}{dr}-\dfrac{i}{r^{2}}\sin\theta\dfrac{\partial}%
{\partial\phi}\right\}  \psi_{\downarrow}\nonumber
\end{align}
is the matrix element of $H_{int}$ Eq.(\ref{h7.01}) between $\Psi_{\downarrow
}$ and $\Psi_{\uparrow}$, and $g(E_{\uparrow})$ is the density of the final
states at energy $E_{\uparrow}$.

The integrand in Eq.(\ref{t2}) consists of a product of three elements: The
function $\psi_{\downarrow}^{\ast}$ whose `width' is about $\Delta
r_{\downarrow}$ (given in Eq.(\ref{down2})) around the origin, An operator
consisting of four $\theta$-dependent terms whose extent around the origin
$\Delta r_{\mu B}$ is roughly $\sqrt{1/K}$ larger than $\Delta r_{\downarrow}$
and the function $\psi_{\uparrow}$ which is an \emph{oscillatory} function
with a characteristic period near the origin $\Delta r_{\uparrow}$ which is
$\sqrt{K}$ \emph{smaller} than $\Delta r_{\downarrow}$. This suggests that we
can approximate the integral in Eq.(\ref{t2}) by substituting $\sin\theta$,
$d\theta/dr$ and $d^{2}\theta/dr^{2}$ by their value at $r=0$,
\begin{align}
\sin\theta &  \simeq\dfrac{B_{\bot}^{\prime}}{B_{0}}r\label{t5}\\
\dfrac{d\theta}{dr}  &  \simeq\dfrac{B_{\bot}^{\prime}}{B_{0}}\nonumber\\
\dfrac{d^{2}\theta}{dr^{2}}  &  \simeq0\nonumber
\end{align}

Substituting Eqs.(\ref{t5}), (\ref{down1}) and (\ref{up0.2}) into
Eq.(\ref{t2}) gives
\begin{equation}
H_{\downarrow,\uparrow}\simeq-\sqrt{\pi}\hbar^{2}\frac{B_{\bot}^{\prime}%
}{mB_{0}}C\sqrt{\dfrac{2}{K}}\exp\left[  -\dfrac{1}{K}\right]  , \label{t8}%
\end{equation}
where we have used the definite integral
\begin{equation}%
{\displaystyle\int\limits_{0}^{+\infty}}
r^{2}J_{1}\left(  br\right)  e^{-ar^{2}}dr=\dfrac{b}{4a^{2}}\exp\left[
-\dfrac{b^{2}}{4a}\right]  . \label{t8.1}%
\end{equation}

When Eq.(\ref{t8}) is substituted into Eq.(\ref{t1}) the term $C^{2}%
g(E_{\uparrow})$ appears. This term can be calculated by temporarily
introducing suitable boundary conditions: Assume that the system is bounded by
an infinite potential wall at $r=R$, the radius $R$ being large compared to
$\Delta r_{\downarrow}$ yet small when compared to $\Delta r_{\mu B}$. In this
case the uniform potential approximation holds for all $r<R$, and the wave
function has the form%
\[
\psi_{\uparrow}\left(  r,\phi\right)  =Cg\left(  r\right)  e^{i\phi/2}\text{,}%
\]
where the radial part $g(r)$ satisfies the Schr\"{o}dinger equation
\[
-\frac{\hbar^{2}}{2m}\left[  \frac{d^{2}g}{dr^{2}}+\frac{1}{r}\frac{dg}%
{dr}-\frac{g}{r^{2}}\right]  =(\mu B_{0}+E)g.
\]
This equation has as non-singular solutions the Bessel functions of order 1,
\[
g(r)=J_{1}(kr)\quad\mathrm{where}\quad k^{2}=\frac{2m}{\hbar^{2}}(E+\mu
B_{0}),
\]
with eigenvalues $k=k_{n}$ determined by the boundary condition. From $R\gg
r_{\downarrow}$ it follows that $kR\gg1$ such that $J_{1}(kR)$ may be
approximated by the first term of its asymptotic expansion. Therefore, the
boundary condition reads
\[
J_{1}(kR)=\sqrt{\frac{2}{\pi kR}}\,\cos\left(  kR-\frac{3\pi}{4}\right)  =0.
\]
This yields the eigenvalues $k_{n}=(n+1/4)\pi/R$. The density of states on the
$k$-axis is thus given by $dN/dk=R/\pi$, from which one obtains the density of
states on the energy axis at $E=\mu B_{0}$
\begin{equation}
\rho(E=\mu B_{0})=\frac{dN}{dE}=\frac{mR}{\pi\hbar^{2}k}=\frac{1}{2\pi}%
\sqrt{\frac{m}{\hbar^{2}\mu B_{0}}}\,R. \label{t8.2}%
\end{equation}
The constant $C$ is determined by the normalization condition
\[
\int|\psi_{\uparrow}|^{2}rdrd\phi=2\pi C^{2}\int_{0}^{R}[J_{1}(kr)]^{2}rdr=1,
\]
which gives\cite{grad}
\[
\int_{0}^{R}[J_{1}(kr)]^{2}rdr=\frac{1}{2}R^{2}[J_{2}(kR)]^{2}.
\]
In the asymptotic region $kR\gg1$, the function $J_{2}(kR)$ takes the values
$\pm\sqrt{2/(\pi kR)}$ at the zeros of $J_{1}(kR)$. This gives
\begin{equation}
C=\frac{k}{2R} \label{t8.3}%
\end{equation}
and therefore
\begin{equation}
C^{2}\rho(E=\mu B_{0})=\frac{m}{2\pi\hbar^{2}}. \label{t13}%
\end{equation}

Finally, using Eqs.(\ref{t13}) and (\ref{t8}) inside Eq.(\ref{t1}) gives
\[
T_{esc}=\dfrac{1}{64\pi\omega_{vib}}\exp\left[  \dfrac{2}{K}\right]
=\dfrac{T_{vib}}{128\pi^{2}}\exp\left[  \dfrac{2}{K}\right]  ,
\]
where $T_{vib}=2\pi/\omega_{vib}$ is the period of classical oscillations
inside the trap.

\section{Discussion.\label{dis}}

Summarizing all we have found we conclude that the problem we have studied has
three important time scales: The shortest time scale is $T_{prec}$, which is
the time required for \emph{one} precession of the spin around the axis of the
local magnetic field. The intermediate time scale is $T_{vib}=T_{prec}/K$,
which is the time required to complete one cycle of the center of mass around
the center of the trap. These two time scales appear both in the classical and
the quantum-mechanical analysis. The longest time scale (provided $K$ is
small) $T_{esc}$, which is not present at the classical problem, is the time
it takes for the particle to escape from the trap.

Whereas the classical analysis yields an upper bound of $K=\sqrt{4/27}$ for
trapping to occur, no such sharp bound exists in the quantum-mechanical
analysis. This is related to the fact that one cannot associate an effective
potential well with a finite barrier with the system.

As an example, we apply our results to the case of a neutron and an atom
trapped with a field $B_{0}=100$ Oe and $B_{0}/B_{\bot}^{\prime}=10$cm. These
parameters correspond to typical traps used in Bose-Einstein condensation
experiments\cite{bec,bec2,bec3,bec4}. The results, being correct to within an
order of magnitude, are outlined in the following table.
\[%
\begin{tabular}
[c]{|c||c|c|}\hline
& \multicolumn{2}{||c|}{$%
\begin{array}
[c]{c}%
B_{0}=100\text{ Oe}\\
B_{0}/B_{\bot}^{\prime}=10\text{cm}%
\end{array}
$}\\\cline{2-3}%
& Neutron & Atom\\\hline
$m$ gr & $\sim10^{-25}$ & $\sim10^{-22}$\\\hline
$\mu$ emu & $\sim10^{-23}$ & $\sim10^{-20}$\\\hline
$K$ & $\sim10^{-5}$ & $\sim10^{-8}$\\\hline
$T_{prec}$ sec & $\sim10^{-6}$ & $\sim10^{-9}$\\\hline
$T_{vib}$ sec & $\sim10^{-1}$ & $\sim10^{-1}$\\\hline
$T_{esc}$ sec & $\sim10^{\left(  10^{5}\right)  }$ & $\sim10^{\left(
10^{8}\right)  }$\\\hline
\end{tabular}
\]

We note that in both cases $K$ is much smaller than $1$. Also, the calculated
lifetime of the particle in the trap is extremely large, suggesting that the
particle (either neutron or atom) is tightly trapped in this field.

The problem studied in this paper deals with a spin $1/2$ particle. Though
this fact has little influence on the solution of the \emph{classical}
problem, the extension to higher spin values complicates the analysis of the
\emph{quantum-mechanical} problem. In this case one has to deal with a
($2S+1$)-component spinor, and the interaction Hamiltonian does no longer
connect the ($-S$)-state to the ($+S$)-state, but only to the ($-S+1$) and
($-S+2$) states which for $S\geq5/2$ will \emph{still} be trapped.

\end{document}